\documentclass[printer]{aa}
\usepackage[varg]{txfonts}

\graphicspath{ {fig/} }

\usepackage{natbib,twoopt}
\usepackage[breaklinks=true]{hyperref}
\bibpunct{(}{)}{;}{a}{}{,} 

\begin{document}

\title{Statistical isotropy violation in WMAP CMB maps resulting from non-circular beams}

\titlerunning{SI violation in WMAP CMB maps due to non-circular beams}
\authorrunning{Das et al.}

\author{Santanu Das
\and Sanjit Mitra
\and Aditya Rotti
\and Nidhi Pant
\and Tarun Souradeep}

\institute{IUCAA, P. O. Bag 4, Ganeshkhind, Pune 411007, India, \email{santanud@iucaa.ernet.in, sanjit@iucaa.in, arotti@fsu.edu, nidhip@iucaa.in, tarun@iucaa.in}}

\date{}

\abstract{
Statistical isotropy (SI) of Cosmic Microwave Background (CMB) fluctuations  is a key observational test to validate   the cosmological 
principle underlying the standard model of cosmology.  While a detection of   SI violation would
have immense cosmological ramification, it is important to recognise their possible origin in systematic effects of observations.
WMAP seven year (WMAP-7) release  claimed significant deviation from SI in the bipolar spherical harmonic (BipoSH) coefficients $A_{ll}^{20}$ and $A_{l-2l}^{20}$.
Here we present the first explicit reproduction of the measurements reported in WMAP-7, confirming that beam systematics alone can  completely account for the measured SI violation. The possibility of such a systematic origin was alluded to in
WMAP-7 paper itself and other authors but not as explicitly so as to account for it accurately.
We simulate CMB maps using the actual WMAP non-circular beams and scanning strategy. Our estimated BipoSH spectra from these maps match the WMAP-7 results very well. 
It is also evident that  only a very careful and adequately detailed modelling, as carried out here, can conclusively establish that the entire signal arises from non-circular beam effect.  This is important since cosmic SI violation signals are expected to be subtle and  dismissing a large SI violation signal as observational artefact  based on simplistic plausibility arguments  run the serious risk of ``throwing the baby out with the bathwater''.}

\maketitle

\section{Introduction}

Cosmic Microwave Background (CMB) measurements have paved the way for the transition of cosmology to a precision science. The recent observations by WMAP and {\it Planck} favour a minimal six parameter $\Lambda$CDM cosmology~\citep{WMAP9-cosmology, planck16}.  An assumption of statistical isotropy (SI) of CMB is a fundamental tenet of the standard model of cosmology, which increasingly precise measurements now enable to be observationally tested. Indeed several SI violation anomalies have been detected in recent measurements of CMB temperature anisotropies~\citep{WMAP9-anomalies, wmap7-anomalies, planck23}. Although there is no dearth of proposed theoretical models~\citep{Ackerman2007, Ramazanov2012, AR-MA-TS-lens-biposh} to explain these SI violations, it is necessary to model all the known systematics and account for the biases introduced by them before probing any deeper for subtle cosmological effects.

It is well known that non-circular instrumental beams can induce artefacts in CMB measurements. The resultant effect on the CMB angular power spectrum ($C_l$) has been extensively studied in literature~\citep{Tegmark2003,Ashdown2007,Fosalba2002,Hinshaw2007,Mitra2009,Mitra2004,
Souradeep2006,Souradeep2001,Das2013a}.  It has been recognised that non-circular beams can lead to SI violation in the observed maps that are not captured in the angular power spectrum~\citep{AH-TS-03,Mitra2004}.  Bipolar spherical harmonic (BipoSH) expansion of the CMB two point correlation function provides a novel tool to probe SI violation in CMB maps~\citep{AH-TS-03}. Detection of non-zero BipoSH spectra, $A_{l_1l_2}^{LM}$ ($L>0$), is the basic probe of SI violation.

 In this work, we address the  high statistical significance detection  of SI violation by the WMAP-7 team  in BipoSH coefficients $A_{ll}^{20}$ and $A_{l-2l}^{20}$ from the WMAP V- and W-band maps~\citep{wmap7-anomalies}. Their paper mentions that ``it seems very likely that the observed quadrupolar effect is the result of incomplete handling of beam asymmetries'', however it was never explicitly shown that the measured BipoSH spectra could be emulated by performing detailed numerical simulations. In a recent paper,  we provided a formalism to  study non-circular beams in BipoSH representation and show that mild non-circularity at levels comparable to WMAP beams do generate significant BipoSH spectra  in observed CMB maps~\citep{Pant2015}. In another approach, \citet{Hanson2010} invoked the anisotropic primordial power spectrum (aPPS) model ~\citep{Pullen2007,Ma2011} and an approximate scan pattern to estimate this effect. However, as shown in \citet{Kumar2015}, it is unlikely that a surrogate estimator can be extrapolated to explain the distinct features observed in BipoSH spectra caused by non-circular beams, namely the observed shift in the location of the primary peak or the zero crossings that are clearly detected in WMAP-7 analysis  of  $A^{20}_{ll}$  in the two bands.
These features arise owing to the combined effect of structures in CMB power spectra and non-trivial beam characteristics~\citep{Joshi2012,Pant2015}, which are discussed in section~\ref{section2}. Given that the detected SI violation signal caused by noncircular beams is very strong, it is important to account accurately for the effect before probing deeper into the residual signature, if any, for cosmological anomalies. Use of an appropriate estimator is thus necessary for the analysis of data from the current and future high-resolution CMB experiments to avoid incorrect or inaccurate prediction of the systematic effects.
  
In this paper, to estimate the full systematic effect accurately, we simulate WMAP observations by incorporating published instrumental beam maps along with the real non-trivial scanning strategy for the V and W bands. BipoSH spectra obtained from these simulated maps match those recovered by an identical analysis of WMAP-7 observed maps, providing the first direct demonstration that the systematic effects that are due to non-circular beams is sufficient to account for the WMAP-7 measurements of quadrupolar anomaly.

\section{Review of BipoSH }

\label{section1}
The standard model of cosmology assumes the universe to be SI. The CMB temperature anisotropies can be expressed in terms
of the spherical harmonics as $T(\hat{n})=\sum_{l=0}^{\infty}\sum_{m=-l}^{l}a_{lm}Y_{lm}(\hat{n})$, where $a_{lm}$ are the coefficients of expansion on this basis. If the CMB temperature anisotropies are described as a Gaussian random field, warranted by recent precise measurements made by {\it Planck}~\citep{planck23,planck24}, specifying the two-point correlation for this field completely characterises its statistical properties. Further, if the basic assumption of SI is valid, then the two-point correlation function is fully described by the equivalent harmonic space quantity, the angular power spectrum $C_{\ell}$ defined as $\left\langle a_{lm}a_{l'm'}^{*}\right\rangle =C_{l}\delta_{ll'}\delta_{mm'}$, where the angular brackets $\langle \cdots \rangle$ denote an average over an ensemble of CMB realizations.

However, in a non-SI universe, the BipoSH spectra capture the full description of  the two-point correlation function~\citep{AH-TS-03}. The BipoSH spectra $A_{ll'}^{LM}$ are defined by the expression~\citep{wmap7-anomalies},
\begin{equation}
\left\langle a_{lm}a_{l'm'}^{*}\right\rangle \ = \ \sum_{LM} A_{ll'}^{LM} \, \mathcal{C}^{LM}_{l m l^\prime m^\prime} \, \mathcal{C}^{L0}_{l 0 l^\prime 0} \, (\Pi_l \Pi_{l'}/\Pi_L),
\end{equation}
where $\Pi_L := (2 L + 1)^{1/2}$ and $\mathcal{C}^{LM}_{l m l^\prime m^\prime} $ are the Clebsch-Gordan coefficients.
While the BipoSH spectrum $A_{ll}^{00} = C_l$ is the standard angular power spectrum, detecting non-vanishing power in the remaining BipoSH spectra $A_{ll'}^{LM},~ L,M \neq 0$ forms the basic criteria for probing SI violation.

\section{Simulation and analysis}

\label{section2}

The WMAP satellite scanned the sky temperature in five different frequency bands at 22, 30, 40, 60, and 90 GHz. The WMAP-7 team searched for SI violation signals in the 60GHz (V) and 90GHz (W) bands, since these channels are the least foreground-contaminated and detected significant non-vanishing BipoSH spectra $A_{ll}^{20}$ and $A_{l-2l}^{20}$. We simulate observed CMB maps, by closely replicating the WMAP instrument characteristics and scanning strategy for  the same bands. 

The V and W band are comprised of two and four detectors respectively. Since each detector has its unique beam pattern and scan path,  it is important to carry out the simulations for each differencing assembly (DA) independently. We generate $30$ realisations of SI CMB skies from the best-fit WMAP-7 angular power spectrum using HEALPix~\citep{HEALPIX}. The procedure that was followed to generate the time order data (TOD) from these realisations and mapmaking is described below.

\subsection{Scan pattern}

Each DA of the WMAP satellite involves a pair of radiometers. The WMAP satellite measures the temperature difference between the DA radiometer pairs. Each of these antennae are inclined at an angle $\sim 70.5^{\circ}$ to the symmetry axis of the satellite. The satellite spins around this axis with $2.2$~min period and slowly precesses about the Sun-Earth axis at an inclination angle of $22.5^{\circ}$, with a precession period $\sim 1$ hour. This particular method of scanning is adapted by the WMAP satellite to reduce noise in the data. This entire system is located at L2 and is moving around the Sun with a period of one year.

We have used an analytical approximation~\citep{Das2013a,moss11} to calculate the pointing of the satellite to obtain the beam locations and their orientations on the sky as a function of time.

\subsection{Beam functions}

The observed temperature $T(\hat{n})$ along a particular direction $\hat{n}$ is related to the underlying sky temperature $\tilde{T}(\hat{n})$ by
\begin{equation}
T(\hat{n}) \ = \ \int B(\hat{n},\hat{n}')\tilde{T}(\hat{n}')d\Omega_{\hat{n}'} \, ,
\end{equation}
 where $B(\hat{n},\hat{n}')$ is the beam response function.
The shape and size of the intrinsic instrumental beams, elliptical Gaussian (EG) profiles fits for FWHM of W (V) band beam are $\sim 13.2'$ ($\sim 21.0'$) with eccentricity $\sim 0.4$ ($0.46$)~\citep{wmap3beam,Mitra2004}. However, EG approximation of the beam profile {does not provide adequately accurate model} for the problem. Specifically, such an approximation cannot explain subtle zero crossing features seen in the BipoSH spectra caused by non-circular beams~\citep{Pant2015}. Moreover, the effect of scanning creates varying effective beam profile across the sky~\citep{febecop}. Therefore, for an accurate estimation of the effect of non-circular beams on the BipoSH spectra, a full convolution of SI CMB sky with the actual beam, along with scanning strategy, is critical.

\begin{figure}[t]
\centering
\includegraphics[width=0.425\textwidth]{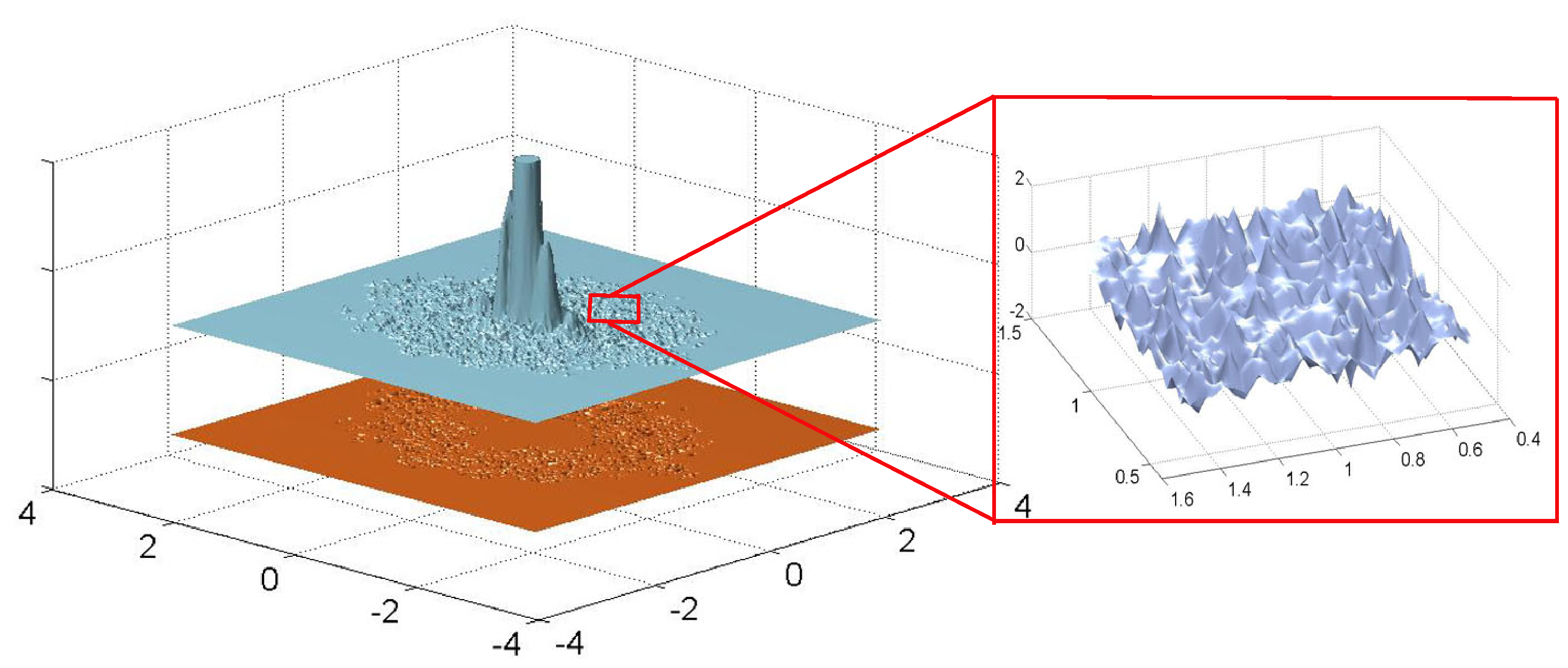}
\caption{\label{fig:beamfunction} Enlarged view of the A-side beam of W3 DA. The central part of the beam is truncated to show the sub-dominant structures in the beam. Apart from the central peak the beam consists of many features extending up to $\sim 4$ degree in an annular ring. The beam also takes unphysical negative values in the annular region, which perhaps arise from noisy measurements. We have separated the negative (orange) and positive (blue) sensitivity parts for visual clarity. A small portion of the beam (with both positives and negatives) is enlarged to show the presence of subtle detailed structures. Motivation for considering all these details in the shape of the beam are described in the text.
}
\end{figure}

WMAP beam maps are available as $600\times600$ arrays on the LAMBDA website\footnote{\url{http://lambda.gsfc.nasa.gov/product/map/dr5/beam_maps_get.cfm}} extending $24^{\circ}$ on each side. Thus each of the pixels covers $2.2'$ of the beam profile. Apart from the central peak, the beam consists of many small structures around the peak, as seen in Fig.~\ref{fig:beamfunction}, which span a much wider solid angle. This includes sub-dominant side peaks and a noisy annular region with negative pixel values, which remain significant up to a radial distance of $\sim 4^{\circ}$ from the central peak. The contribution from the outer region of the beam is critical for reproducing the amplitude and features of the observed BipoSH spectra, most importantly when explaining the flipping of signs of BipoSH spectra seen in WMAP-7 year observations.
We note that the negative pixels of the beam sum up to $\sim 10\%$ of the total integrated beam profile.
For completeness, in our analysis we have taken a $\pm 6^{\circ}$ cut-off for the beam around the pointing direction, while numerically convolving it with the sky map using Eq.~(\ref{eq:Beam_Convolution}).
We also tested that ignoring the part beyond $6^{\circ}$ does not affect our results.

\subsection{Generation of time-ordered data and map-making}

The observed temperature anisotropy along a given pixel direction $\hat{n}$ can be expressed as a discretised convolution as
\begin{equation}
T(\hat{n}) \ = \ \sum_{i\in S_{\hat{n}}}B(\hat{n}'_i,\hat{n})\tilde{T}(\hat{n}'_i) \, \bigg/ \, \sum_{i\in S_{\hat{n}}}B(\hat{n}'_i,\hat{n}) \, ,\label{eq:Beam_Convolution}
\end{equation}
 where $\hat{n}$ is a pointing direction, $\hat{n}'_i$ is the central direction of the $i^\text{th}$ pixel and $S_{\hat{n}}$ is the set of all the pixels for which $B(\hat{n}'_{i},\hat{n})$ is non-zero. If the beam function were considered to be elliptical Gaussian, pixels within $3$-$4\sigma$ would have been sufficient for the numerical integration. However, for the actual WMAP beams, as shown in Fig. \ref{fig:beamfunction}, where we plotted the beam map of the A side of W3 DA, this area cut-off is not valid and requires integration over the entire beam. 

To compute $T(\hat{n})$, we include all the pixels inside the radius $\pm6^{\circ}$, from that pointing direction and then interpolate the beam function at the centre of the pixels for convolution using Eq.(\ref{eq:Beam_Convolution}). The WMAP team provides the temperature sky maps at $N_{side}=512$. At this resolution, the distance between two adjacent pixel centres is $~3.8'$. As the sizes of many of the noisy peaks in the skirt of the beams are smaller than $3.8'$, these features will be missed in numerical integration, hence the convolution is not precise.
To overcome this problem, all the simulations in the paper are generated at a resolution $N_{side}=1024$, for an accurate evaluation of the numerical beam convolution of the map, following which the pointing direction is converted to the pixel numbers that correspond to a $N_{side}=512$ resolution map.

To evaluate the beam function at the centre of each pixel, it is necessary to use a fast and accurate interpolation scheme. Linear interpolation at the centre of each pixel by considering the value of the beam function at the four pixels around it only works for the central peak. However, the features in the annular region that surround the central peak of the beams being of a rapidly varying nature,  the error due to linear interpolation is found to be unacceptably large.  In this region, a non-linear interpolation scheme, such as the cubic or the spline interpolation, may provide reasonable results. But these methods are computationally expensive. To overcome this issue, we upgraded the beam function from $600\times600$ grid to $2400\times2400$ grid using MATLAB's {\tt spline2} interpolation scheme. This high-resolution grid enables us to use a linear interpolation of the beam in the convolution process without compromising the precision and is computationally fast. We also checked that a further increment in the interpolated grid size is not required to improve accuracy.

We follow this procedure to generate noiseless TOD, where each sample is obtained by Eq.~(\ref{eq:Beam_Convolution}). The TOD vector for each DA can then be written as $d \ = \ A \, T$, where $T$ is the scanned sky temperature and $A$ is the pointing matrix, where each row consists of two non-zero components, one $+1$ and one $-1$. We use the Jacobi iteration method~\citep{Das2013a} for map-making, although other possible methods are discussed in~\citet{Hamilton2003}.

\subsection{BipoSH analysis in the presence of masks}

The observed CMB maps are contaminated by foregrounds. Even though the W and V band maps are the least foreground contaminated, the region close to the galactic equator has to be masked to extract information on the genuine CMB signal. The application of masks in analyses as well as the presence of anisotropic noise in the maps, correlates different harmonic modes of CMB anisotropies, resulting in non-vanishing BipoSH spectra ($L \ne 0$). To remove this bias from observed WMAP maps, masked with KQ-75, firstly we evaluate BipoSH coefficients from $1000$ masked realisations generated from a {\tt synfast} subroutine of HEALPix for the best-fit CMB power spectrum and average beam transfer function. This estimate is used to debias and arrive at the final WMAP BipoSH spectra. The results obtained closely match the WMAP findings.

We now apply our WMAP analysis identically to our simulated maps with a non-circular beam and scan strategy. Even though we do not include foregrounds in our simulations, we use the mask in our analysis to compare them with the results derived from the observed maps.

\section{Results}
\label{section3} 

\begin{figure}[t]
\centering
\includegraphics[width=0.4\textwidth]{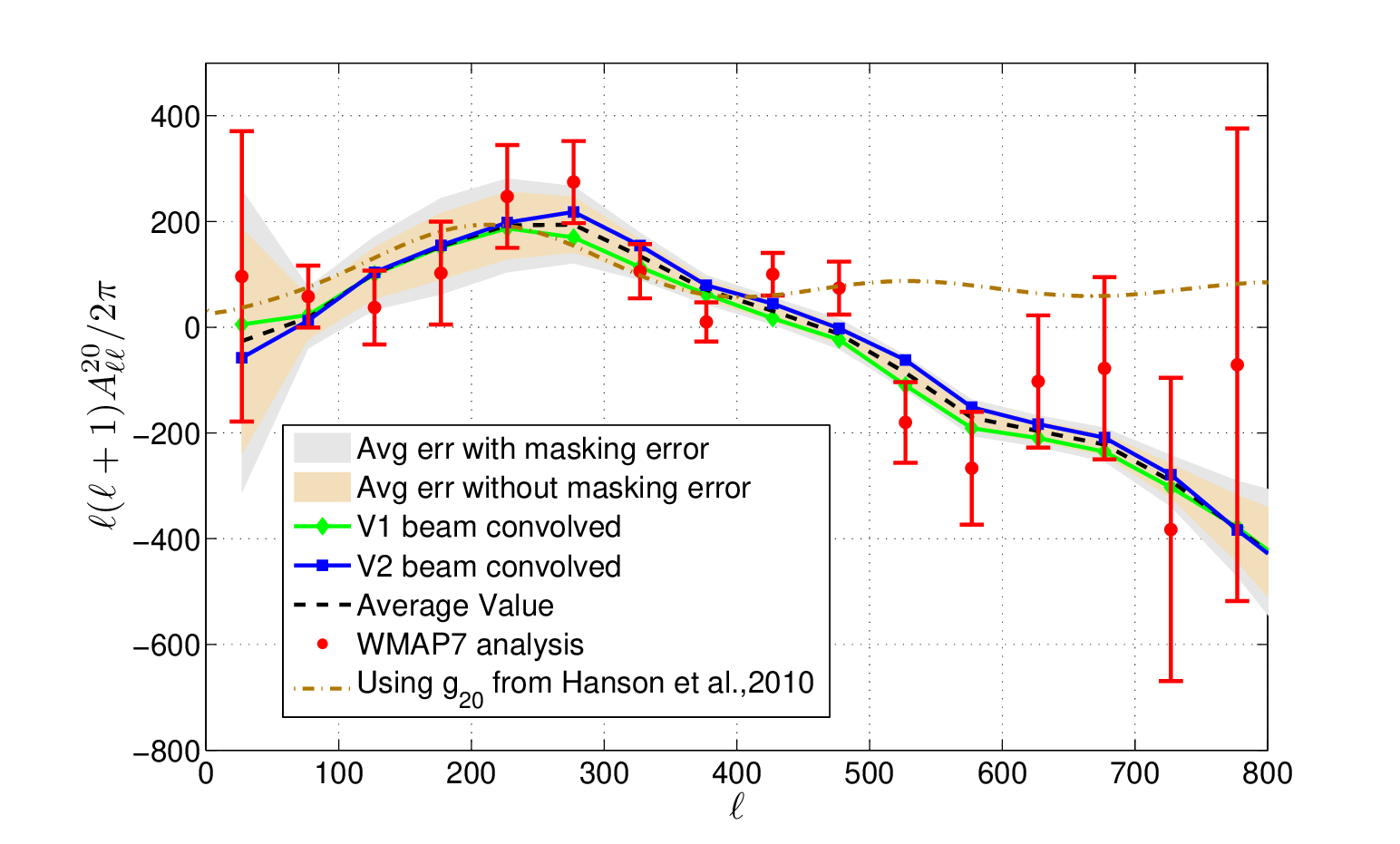}
\includegraphics[width=0.4\textwidth]{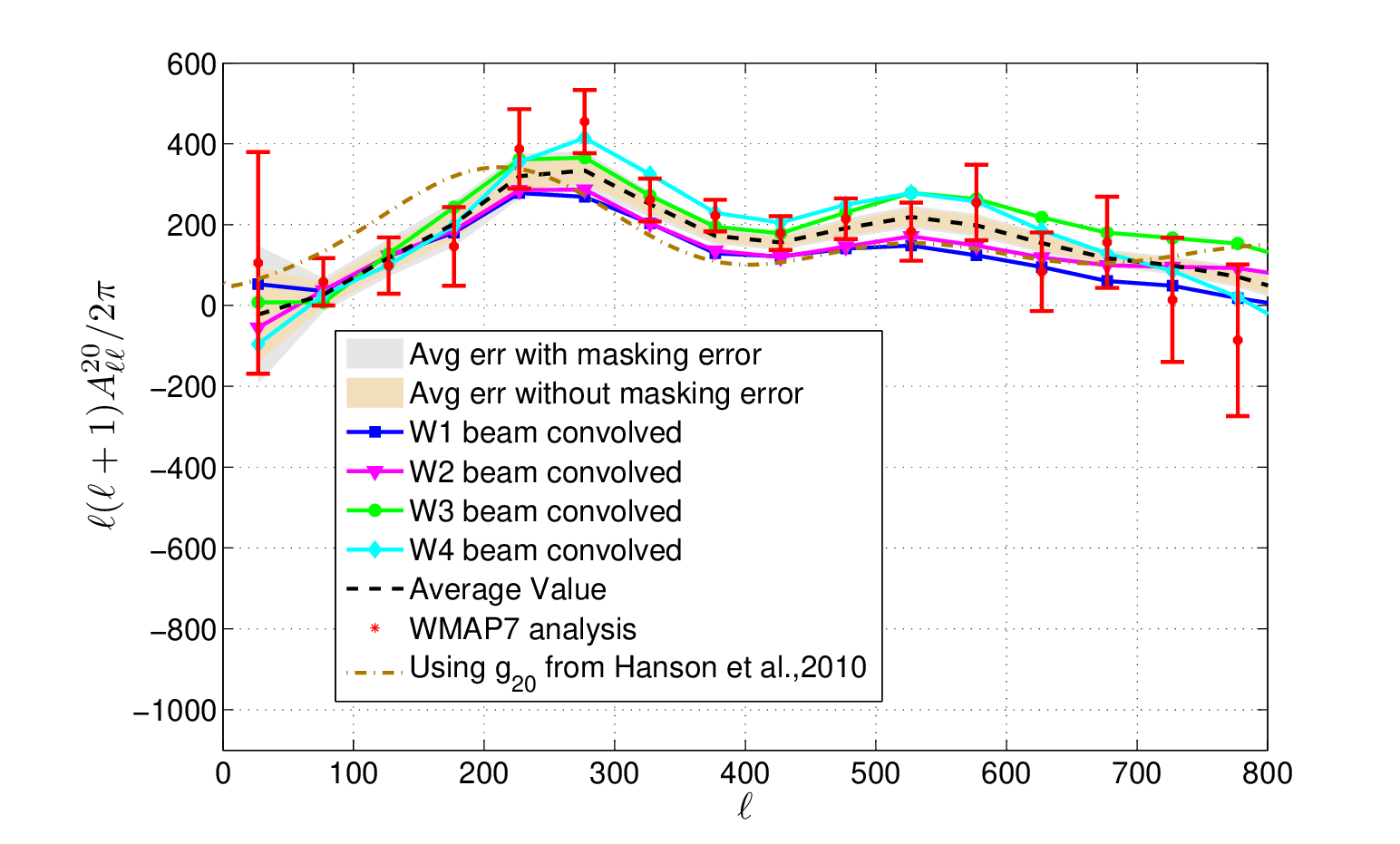}
\caption{\label{fig:biposh} Comparison between the beam window function corrected BipoSH spectra $A_{ll}^{20}$ obtained from WMAP maps (red errorbars), observed in V (top) and W (bottom) bands, and our detailed simulations of the respective channels (solid lines). The average BipoSH spectra across the DAs are denoted by the dashed lines. The saffron band shows the standard error derived from all the simulations for each band only (total $60$ simulations for V and $120$ for W), while the grey band also combines the error on the mask bias correction.
Spectra from the simulated maps match the observed spectra very well, correctly reproducing the location of the peak and the zero crossings.
%
Also plotted are the spectra (dash-dot lines) evaluated from the aPPS model parameterised by $g_{20}$ provided in Table~1 of \citet{Hanson2010} to emphasise that a surrogate estimator is unlikely to capture these subtle effects.
}
\end{figure}

The WMAP-7 team detected an SI violation signal in BipoSH spectra in both the V and the W bands.  The primary goal of this paper is to make a detailed computation of the BipoSH spectra arising from the actual WMAP beam, coupled to  the scan strategy, and establish conclusively that the measured non-zero $L=2$ BipoSH spectra in WMAP-7 can be entirely attributed to this systematic effect. We generate $30$ simulations each for four DAs in the W band (W1, W2, W3, W4) and two DAs in the V band (V1, V2) by convolving CMB sky realisations with the actual WMAP beam and scanning strategy and performed BipoSH analysis on each set separately. The BipoSH spectra obtained from the true and simulated datasets are shown in Fig.~\ref{fig:biposh}.

In Fig.~\ref{fig:biposh}, we plot the WMAP data points (red error bars). The average $A_{ll'}^{LM}$s from the four DAs are plotted in black dashed lines.
%
%
The location of peak of the $A_{ll}^{20}$ spectra is set by both the first CMB acoustic peak, and the  parameter $b_{l2}$ characterising beam noncircularity~\citep{Pant2015}.
A closer inspection of the $A_{ll}^{20}$ spectrum reveals a change in sign, for both bands, which can only be explained by taking cognisance of a large noisy region around the main peak in the WMAP beams shown in Fig.~\ref{fig:beamfunction}.
Thus the spectra from the simulated maps match the observed spectra very well, replicating the non-trivial features, leaving very little scope for accommodating additional systematic or cosmological effects.
%
%
The error bars are estimated from $1000$ simulations used for estimating the mask bias after adding anisotropic noise~\citep{AR-MA-TS-lens-biposh} and scaled for different cases based on the total number of simulations. The error bars are not plotted for each individual DA to avoid clutter in the plot, rather we  plotted the error bars of the average spectra. We plotted two error bars, the regions shaded in orange denote standard errors estimated from the DA simulations alone, while in the grey region we also include the standard errors on the estimated mask bias from $1000$ simulations in quadrature. The large errorbars seen at high-$l$ in the WMAP analysis is mostly due to noise in the data.
Since we plot errorbars on spectra averaged over many simulations, they are smaller (at least by a factor of few).
The non-zero measurements of $A_{ll-2}^{20}$ spectrum reported in WMAP-7 (not plotted here) can also be convincingly reproduced by our simulations.
For the W band,
it can be seen in the figure that
BipoSH spectra for W1 and W2 are very close to each other and similarly for W3 and W4. This trend is essentially due to similarities in their respective beam transfer functions.

To highlight the reliability of our method and to emphasise the use of a proper estimator for SI violation, we plot the $A_{ll}^{20}$ for V and W bands derived from the aPPS model-based estimates~\citep{Hanson2010} in Fig.~\ref{fig:biposh}. This model predicts $A_{ll}^{20} = g_{20} \, C_\ell / \sqrt{4\pi}$, where $g_{20}$ lacks a multipole dependence that is crucial for characterising the beam's noncircularity, hence fails in recovering the shift in the location of the peaks and the zero crossings of the BipoSH spectra. The match at the first peak in $A_{ll}^{20}$ for the V band is due to a mere coincidence that both $b_{l2}$ and $C_l$ peak around $l \approx 220$, which is not the case for the W band, where $b_{l2}$ peaks near $l \approx 350$~[See Fig.5 \citep{Pant2015}].

\begin{figure}[h]
\centering
\includegraphics[width=0.4\textwidth]{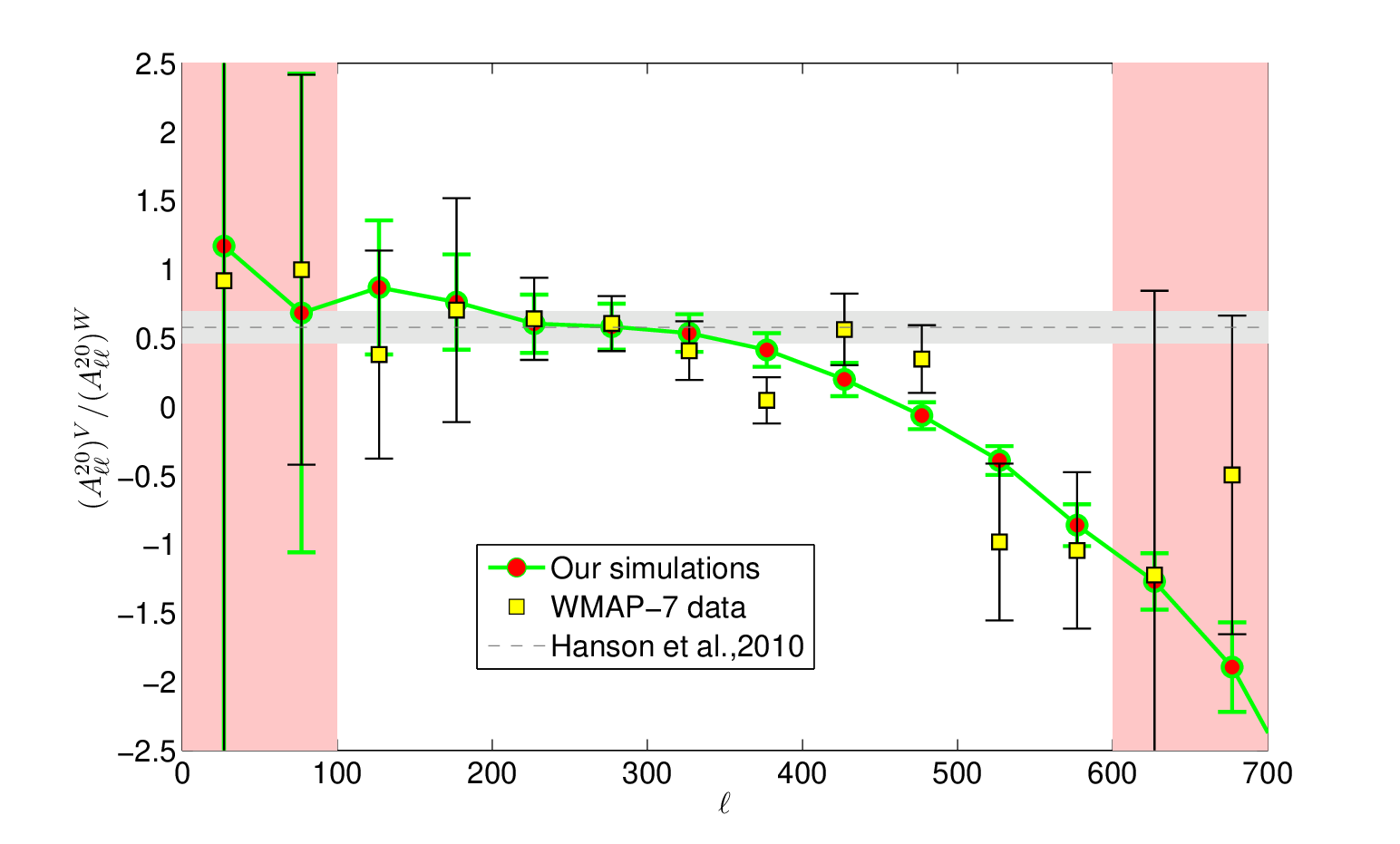}
\caption{\label{fig:ratio} Ratio of the BipoSH spectra $A_{ll}^{20}$ for the V and the W bands. The trend recovered by our detailed analysis clearly matches the WMAP7 observations. The very low and high multipole regions which have poor S/N are indicated with pink shading.
We also plot the ratio corresponding to \citet{Hanson2010} estimates, which is consistent with observations at low-$\ell$s, but, not surprisingly, far off at high-$\ell$s.}
\end{figure}

We also plot the ratios of observed and predicted $A_{ll}^{20}$ for V and W bands for our method as well as the same derived from the aPPS model-based estimates in Fig.~\ref{fig:ratio}.
Our method clearly produces a close match with observation, while it is clear that the aPPS estimator is not designed to capture the observed trend.

\section{Discussion and conclusion}
\label{section4}

In this paper, we have demonstrated for the first time that the quadrupole anomaly measured in WMAP-7 maps can be closely reproduced by incorporating the effect of non-circular beams and scan strategy of the satellite. This clearly indicates, as was also suspected by the WMAP team, that this signal does not have any cosmological origin. The absence of similar detections in BipoSH spectra $A_{ll}^{20}$ and $A_{l-2l}^{20}$ in the more recent measurements by the {\it Planck} satellite~\citep{planck23} reinforces this claim, where an analysis strategy similar to the one described in this letter was employed to account for the non-circular beams. Prior to that, the WMAP nine year analysis team released a set of beam-symmetrised (deconvoled) maps, which also did not show this particular SI violation signature~\citep{Ramazanov2013}. However,  as cautioned by the WMAP team, the deconvolution procedure renders the resultant maps unsuitable for cosmology analyses. Also,  new SI violation signals are present in those maps at high-$l$~\citep{WMAP9-anomalies}.
However, in our so-called forward, approach we evade the complications of deconvolution by the ability to  estimate a correctable bias.

The close match we obtain clearly shows that finer details like foreground residuals, coupling between beam and mask, weak lensing and anisotropic noise, etc., which,  in principle, could have been important for these BipoSH spectra, cause substantially weaker SI violation in WMAP-7 relative to the non-circular beam. This provides a reliable route to probing SI violation from the standard model of cosmology, which will be targeted by the current and upcoming CMB missions.

\begin{acknowledgements}
We would like to thank Saurabh Kumar, Gary Hinshaw, and Eiichiro Komatsu for useful comments and help. SD and AR acknowledge the CSIR, India for financial support through Senior Research Fellowships. SM acknowledges the Fast Track grant SR/FTP/PS-030/2012 of SERB, India. TS acknowledges Swarnajayanti fellowship grant of DST India. Computations were carried out at the HPC facilities at IUCAA.
\end{acknowledgements}

\bibliographystyle{aa}
\bibliography{BBSH.bib}

\end{document}